\documentclass[fleqn,11pt]{olplainarticle}
\usepackage[colorlinks=true, citecolor=red]{hyperref}
\usepackage{semantic}
\usepackage{mathtools}

\title{A Brief Survey of Formal Models of Concurrency}

\author[1]{Charles Averill}
\affil[1]{charles@utdallas.edu}


\newcommand{\picalc}[0]{$\pi$-calculus}

\newcommand{\triple}[3]{#1 \{#2\} #3}

\newcommand{\orline}{\mathrel{\underline{\text{\texttt{or}}}}}

\newcommand{\mea}[0]{Milner et al.}

\newcommand{\infrule}[3]{
\begin{align*}
    \inference{#1}{#2}[\textsc{#3}]
\end{align*}
}

\begin{abstract}
The ubiquity of networking infrastructure in modern life necessitates scrutiny into networking fundamentals to ensure the safety and security of that infrastructure.
The formalization of concurrent algorithms, a cornerstone of networking, is a longstanding area of research in which models and frameworks describing distributed systems are established.
Despite its long history of study, the challenge of concisely representing and verifying concurrent algorithms remains unresolved. 
Existing formalisms, while powerful, often fail to capture the dynamic nature of real-world concurrency in a manner that is both comprehensive and scalable.
This paper explores the evolution of formal models of concurrency over time, investigating their generality and utility for reasoning about real-world networking programs.
Four foundational papers on formal concurrency are considered: Hoare's \emph{Parallel programming: An axiomatic approach}~\cite{hoare}, Milner's \emph{A Calculus of Mobile Processes}~\cite{milner}, O'Hearn's \emph{Resources, Concurrency and Local Reasoning}~\cite{ohearn}, and the recent development of Coq's Iris framework~\cite{iris}.
\end{abstract}

\begin{document}

\flushbottom
\maketitle

\section*{Introduction}


Networks are a cornerstone of modern critical infrastructure, supporting everything from communication systems to power grids and healthcare services.
As these systems are integral to the functioning of society, their safety and security are of paramount importance.
To ensure that these systems are secure and reliable, they require rigorous scrutiny and careful verification.

However, traditional testing and analysis methods often fall short of providing the absolute certainty needed for such critical systems. 
To address this, formal models and verification methods are necessary. These models allow us to represent complex systems in a way that enables structured and exhaustive reasoning about their properties, ensuring that they operate safely and securely.

A network is inherently a complex system, with many sub-areas that require formalization.
Among the most fundamental aspects of networks is concurrency. 
Concurrency describes the simultaneous execution of multiple processes or threads, a concept that underpins nearly all networking software. 
Whether it's handling multiple communication protocols, managing parallel data streams, or synchronizing processes, concurrency plays a vital role in ensuring the functionality of network systems.

As such, formalizing concurrency is not only crucial for understanding the behavior of networking software but also essential for ensuring its safety and security. 
By rigorously proving that concurrent systems are safe, we can provide the confidence needed to trust critical network infrastructures in both everyday and high-stakes contexts.
This paper explores various means of formalizing concurrency in the name of enabling rigorous proofs of concurrency-safety, ultimately supporting the security and reliability of our critical infrastructure.

\section{Parallel programming: An axiomatic approach}

\subsection{Background}

This work is a direct extension of \emph{Hoare logic}, a comprehensive system of reasoning about arbitrary imperative programs developed four years earlier by Hoare~\cite{hoarelogic}.
The key developments of the logic are the \emph{Hoare triple} and the \emph{inference rules}.
A Hoare triple, denoted $\triple{P}{c}{Q}$, states ``if statement $P$ is known to be true and program $c$ is executed, statement $Q$ is true.''
An inference rule simply organizes preconditions and postconditions, so $\frac{P\ \ Q}{R}$ states that $R$ is true if both $P$ and $Q$ are true.

\begin{figure}
    \centering
    \mbox{\inference{}{\triple{P}{\texttt{skip}}{P}}}
    \quad
    \mbox{\inference{}{\triple{P[e/x]}{x \texttt{ := } e}{P}}}
    \quad
    \mbox{\inference{\triple{P}{c_1}{Q} & \triple{Q}{c_2}{R}}{\triple{P}{c_1 \texttt{;} c_2}{R}}}
    \quad
    \mbox{\inference{\triple{P \land b}{T}{R} & \triple{P \land \neg b}{F}{R}}{\triple{P}{\texttt{if } b \texttt{ then } T \texttt{ else } F}{R}}}

    \vspace{2em}
    
    \mbox{\inference{P_2 \Rightarrow P_1 & \triple{P_2}{c}{Q_2} & Q_2 \Rightarrow Q_1}{\triple{P_1}{c}{Q_1}}}
    \quad
    \mbox{\inference{\triple{P \land b}{c}{P}}{\triple{P}{\texttt{while } b \texttt{ do } c}{P \land \neg b}}}

    \caption{Inference Rules in Hoare logic}
    \label{fig:hoarerules}
\end{figure}

Hoare presents the inference rules for a simple, Turing-complete imperative language, shown in Figure~\ref{fig:hoarerules}.
This set of rules allows for a programmer to rigorously prove that code meets a provided mathematical specification.
However, the target language is sufficiently restricted that describing concurrent code with this system alone would be inefficient and difficult to use, hence this work's extension to target parallelism.


\subsection{Disjoint Processes}

Hoare begins by describing the circumstances of executing two \emph{disjoint} processes $Q_1$ and $Q_2$ in parallel.
Two processes are disjoint if they never interact, either by communication over a shared channel or by modifications to shared memory (both generalizations of the same concept).

We are introduced to the first axiom of the extended Hoare logic proposed:
\infrule{\textsc{Disjoint}(Q_1, Q_2) & \triple{P}{Q_1}{S} & \triple{S}{Q_2}{R}}{\triple{P}{Q_1 // Q_2}{R}}{(Asymmetric Parallel Rule)}

This rule states that for any two disjoint processes $Q_1$ and $Q_2$: if you execute $Q_1$ and $Q_2$ in parallel whenever you know some $P$, you know $R$ upon both threads' termination if $Q_1$ and $Q_2$ are disjoint, and if executing $Q_1$ when you know $P$ means you know $S$, and if executing $Q_2$ when you know $S$ means you know $R$.
Hoare aims to show that parallel execution of disjoint processes is functionally equivalent to executing each process in sequence, or,
\begin{align*}
    Q_1 // Q_2 \equiv Q_1; Q_2.
\end{align*}

This is qualitatively proven by decomposing each process into a commutative set of \emph{units}, or individual steps taken by a program.
Hoare considers the unit assignments $x_1 \coloneqq e_1$ and $x_2 \coloneqq e_2$ as an example. Because $Q_1$ and $Q_2$ modify no shared memory, $x_1$ and $x_2$ can be assigned in any order, and $e_1$ and $e_2$ will maintain their value regardless of the order of the assignments:
\begin{align*}
    (x_1 \coloneqq e_1; x_2 \coloneqq e_2) \equiv (x_2 \coloneqq e_2; x_1 \coloneqq e_2).
\end{align*}

This argument can be similarly repeated for each of the remaining constructs of Hoare's language, showing that parallel execution of disjoint processes is indeed equivalent to their sequential execution.
Incidentally, this argument also shows the commutativity of the parallel operator $//$ for disjoint processes, as noted by Hoare later in this work.

\subsection{Competing Processes}

\emph{Competing} processes are those that require access to limited resources such as memory or device access.
Hoare comments that the problem of reasoning about programs laying claim to resources has already been tackled by programming languages with the \emph{declaration}, a standard language construct with well-behaved properties.
The treatment of declarations (and especially the liveness of the declared resource) as a solution to competition over limited resources is informal, but clearly foreshadows the concepts of semaphores and mutual exclusion.

\subsection{Cooperating Processes}

\emph{Cooperating} processes update shared memory via commutative units.
Hoare points out that the treatment of disjointedness and competition allows for automatic, compiler-generated guarantees about safety with the aid of explicit language constructs.
However, cooperation requires the programmer to manually verify that their units of computation properly commute in order to rely on any axiom of parallel execution.

When discussing cooperation, Hoare details that relying on unit computations is largely a pleasant abstract property and has to be viewed critically on a physical machine.
The primary complication approached is that high-level unit computations on real machines might not actually be units, but may take many execution steps to resolve, opening the possibility of race conditions.
In this case, Hoare prescribes the use of exclusion principles to prevent different processes from clobbering each others' data.

\subsection{Communicating Processes}

Hoare details a problem with the previously-introduced rules for parallel computation: each prohibits direct communication between processes by requiring commutativity of units.
To solve this, the notation of \emph{semicommutativity} is introduced.
We say that units $q_1$ and $q_2$ of processes $Q_1$ and $Q_2$ semicommute if \[q_2; q_1 \sqsubseteq q_1; q_2,\]
or that the two permutations have identical effects under all circumstances on all program variables, provided that $q_2; q_1$ terminates. 
Hoare uses this relationship to define communication: if all units $q_1$ and $q_2$ semicommute, then $Q_1$ and $Q_2$ are communicating processes, where $Q_1$ is the producer process and $Q_2$ is the consumer process.
Communicative actions are further defined as semicommutative pushes and pops onto an abstract, shared data structure.

We are finally taken to the updated axiom for communication, encoding semicommutativity:
\infrule{\triple{P_1 \land S_1}{Q_1}{S_1 \land R_1} & \triple{P_2 \land S_1}{Q_2}{S_2 \land R_2}}{\triple{P_1 \land P_2}{Q_1 // Q_2}{R_1 \land R_2}}{(Rule of Two-Way Communication)}

\subsection{Colluding Processes}

Finally, Hoare presents the interesting case of \emph{collusion}, when multiple processes are dispatched to solve a problem in different ways.
These processes can stop early if they discover their approach is insufficient to solve a goal.
The rule provided is
\infrule{\triple{P_1}{Q_1}{R_1} & \triple{P_2}{Q_2}{R_2}}{\triple{P_1 \land P_2}{Q_1 \orline Q_2}{R_1 \lor R_2}}{,}
where $Q_1 \orline Q_2$ denotes that at least one of $Q_1$ or $Q_2$ terminates, but which process terminates is not important.

\subsection{Summary and Analysis}

This work presents a formal basis for the concepts of parallel programming by directly extending the system of Hoare logic.
The extended Hoare logic is still limited, but is adequate for reasoning about the behavior of simple concurrent systems with minimal resource sharing.
Multiple special cases of parallelism are presented alongside an in-depth discussion of the challenges of reasoning about such cases.
Hoare develops a number of axioms for parallel reasoning that can be implemented directly into compilers, as well as axioms that are intended to be used in manual reasoning about the execution of these programs.
Hoare correctly predicts the necessity of semaphores for reasonable concurrent execution, and regularly roots his arguments in real-world examples and issues.

This work is largely qualitative, failing to explain why the provided axiomatic basis is sufficient for all parallel computation, and excluding some thorough explanation of axioms carried over from other works.
Despite these shortcomings, it is clear that Hoare's baseline for parallelism was foundational to the field, given its many correct predictions and enduring paradigms.

\section{A Calculus of Mobile Processes}

\subsection{Background}

This work presents a novel \emph{process calculus}, a low-level system that describes the behavior of concurrent programs.
The family of process calculi are a descendant of the more familiar family of $\lambda$-calculi, which describe the behavior of arbitrary, typically single-threaded programs.
Such a calculus is intended to serve as the formal foundation for the development of more complex languages.
Compilers and interpreters will often convert programs from a high-level language into a low-level calculus or imperative computation model in order to provide guarantees about optimization, safety, and correctness among other properties.

The \picalc\ is not the first process calculus, but instead a generalization and development on others, most directly Milner's Calculus of Communicating Systems~\cite{ccs}.
Other significant process calculi include Hoare's Communicating Sequential Processes~\cite{csp} and Bergstra and Klop's Algebra of Communicating Processes~\cite{acp}.


\subsection{Syntax and Semantics}

\begin{figure}
    \centering
    $P \coloneqq\ 0 \mid P_1 + P_2 \mid \overline{y}x.P \mid y(x).P \mid \tau . P \mid (P_1 | P_2) \mid (x)P \mid [x = y]P \mid A(y_1, \dots y_n)$
    \caption{Syntax of the \picalc}
    \label{fig:picalcsyntax}
\end{figure}

Similar to other process calculi, the \picalc\ as proposed by \mea\ defines the \emph{process}, or agent, as the fundamental unit of a program.
Alongside this fundamental definition is that of \emph{names}, arbitrary identifiers that constitute both data and the names of channels by which processes can communicate.

The \picalc\ specifies a number of process operators for concurrency and communication, as shown in Figure~\ref{fig:picalcsyntax}.
In order, we are given (1) the process that does nothing (2) the process that behaves either like $P_1$ or $P_2$ (3) the process that outputs $x$ on port $y$ and then executes $P$ (4) the process that binds a value $x$ read from channel $y$ and then executes $P$ (5) the process that performs the ``silent action'' and then executes $P$ (6) the process that executes $P_1$ and $P_2$ in parallel (7) the process that defines $x$ to be a private channel among all members of $P$, and executes $P$ (8) the process that executes $P$ if names $x$ and $y$ are equal, otherwise behaves like $0$, and (9) the process that is described by an equation of substitutions.

As an example, consider the following evolution of a \picalc\ system:
\begin{align*}
    \overline{y}m.0 \mid y(z).z(x).0 \longrightarrow 0 & \mid m(x).0
\end{align*}
where the name $m$ was transmitted from process 1 to process 2 via port $y$.
In the subsequent evolution step, we can see that occurrences of $z$ have been substituted with $m$.

The authors place great significance on the distinction between the functionality of names in the $\pi$- and $\lambda$-calculi.
In the $\lambda$-calculus, names are treated as a kind of meta-object, acting as placeholders for expressions in the language that can be substituted in at will.
However, in the \picalc\ names may only be instantiated to other names --- they exist as a distinct class of object within the language that cannot be used to represent agents or expressions.
The authors make this choice, treating names and agents as independent objects, deliberately to avoid passing agents as a value over channels.
Allowing passing agents by value would result in:
\begin{itemize}
    \item Unnecessary or unwanted replication of agents
    \item Compulsory intrusion of one process into the internals of another, possibly revealing secret details of communications with another process
    \item The failure of the \picalc\ to succinctly represent the real-world scenario of transmission of access links.
\end{itemize}

Additionally, the authors point out that \emph{free names}, names which are not bound via an action such as $\overline{y}(x).P$, effectively function as knowledge of the linkage between agents.
This treatment of names is extremely powerful, allowing for the trivial encoding of primitives into the language while generalizing name behavior described in other process calculi such as CCS.

The authors go on to specify a number of process constructs that can be used to solve some standard problems in concurrent systems.
The \emph{executor} receives a link $y$ on channel $x$ and then activates that link.
This typically acts as a trigger to call upon another process $P$ to begin execution:
\begin{gather*}
    Exec(x) \coloneqq x(y).\overline{y}.0 \\
    (\overline{x}z.z(\varepsilon).P \mid Exec(x)) \longrightarrow (\overline{x}z.z(\varepsilon).P \mid x(y).\overline{y}.0) \longrightarrow (z(\varepsilon).P \mid \overline{z}.0) \longrightarrow (P \mid 0)
\end{gather*}
The \emph{copyer} is a construct that, when treating names as constant values such as booleans, can be used to copy a value from one process to another.
$Copy$ turns out to be useful for defining arbitrary operations on these values as well.
The following definition is specified for booleans but can be trivially expanded for any finite set.
\begin{gather*}
    Copy(y, z) \coloneqq ([y = T]\overline{z}T.0) + ([y = F]\overline{z}F.0) \equiv y:[T \Rightarrow \overline{z}T.0, F \Rightarrow \overline{z}F.0] \\
    And(x, y, out) \coloneqq x:[T \Rightarrow Copy(y, out), F \rightarrow \overline{z}F.0]
\end{gather*}
\mea\ continue with an embedding of the combinator and $\lambda$ calculi into the \picalc, displaying its ability to represent arbitrary systems.

\subsection{Scope Extrusion}

A key development of the \picalc\ is that of \emph{scope extrusion}, or expanding the scope of a name from a set of processes to a larger set of processes.
Scope extrusion provides the \picalc\ with the power to represent evolving communication infrastructure, as private links are authenticated and de-authenticated between various parties.
This representation power is critical for reasoning about real-world concurrent systems such as the internet, in which communication privacy is dictated by chaotic and complex authentication schemes.
Additionally, the generalization of scope extrusion to ``pass by reference'' schemes provides the \picalc\ with the ability to model arbitrary systems of computation such as the $\lambda$-calculus.

\subsection{Bisimilarity}

Another key development of the \picalc\ is its \emph{strong bisimilarity}.
\mea\ reintroduce the definitions for simulation relations and emphasize that one of their primary contributions is to ``develop the properties of strong bisimulation and equivalence.''
This development of bisimilarity is motivated by the need to formally describe the relationship between the behaviors of concurrent systems.
The need for a special definition for what would seem like a standard functional correctness principle arises because of the tendency for concurrent systems to be designed not to terminate.
Two processes are said to be bisimilar ($P \dot\sim Q$) if: $P \longrightarrow P'$ implies that there exists $Q$, $Q'$ such that $Q \longrightarrow Q'$ and $P' \dot\sim Q'$.
Intuitively, these processes are bisimilar if they act identically to an outside observer after some number of program steps.

The authors present a number of properties of bisimilarity, notably
\begin{itemize}
    \item Congruence (an equivalence relation on processes) is a strong bisimulation
    \item Strong bisimilarity is an equivalence relation
    \item There exist multiple simple bisimulations for common processes, such as
    \begin{gather*}
        P + P \dot\sim P \qquad P_1 | P_2 \dot\sim P_2 | P_1 \qquad (y)(P_1 \mid P_2) \dot\sim (y)P_1 \mid (y) P_2 \textrm{ if } y \notin freenames(P_1 \mid P_2) \qquad [x = x]P \dot\sim P
    \end{gather*}
\end{itemize}

\subsection{Summary and Analysis}

This work presents a basis for all concurrent programs by encoding fundamental properties of concurrency into a process calculus.
The semantics of this language are explored in detail alongside a number of examples that present real-world issues in networking in terms of the \picalc.
Additionally, a known method for establishing correctness of non-terminating programs is expanded upon in great detail and rigor, establishing a number of rules to be used for reasoning about these programs at a much higher level than would be possible otherwise.

\mea\ develop a comprehensive foundation for concurrent programs sufficiently succinct to reasonably encode all classical networking scenarios.
Its presentation is notably formal but frequently calls back to practical applications to ground the work in real-world systems, providing a healthy balance of detail for implementation of the calculus.

\section{Resources, Concurrency and Local Reasoning}

\subsection{Background}


This work directly extends \emph{Separation Logic}, a direct extension of the aforementioned Hoare logic.
Separation Logic intends to reason about the processes accessed and modified by programs as the state of the machine evolves.
A core concept is the \emph{heap}, an abstract representation of a memory state.
Heaps can be combined to make larger heaps.
Another foundation of the logic is the \emph{frame rule}, an axiom that describes the ability to partition heaps into groups of resources, allowing for proofs describing the behavior of programs which only have access to specific heaps.
As O'Hearn details, the frame rule turns out to generalize surprisingly well to shared resources among concurrent processes.

An important concept touched on many times in this work is that of the \emph{semaphore}, and the related \emph{mutex} or \emph{mutual exclusion} group.
A semaphore acts as a counter that signals when a specified resource is available for access and modification.
Semaphores can define how many threads can gain access to their resources, as a program will request and return access to the resource via guards.
A mutex is a specific case of a semaphore, where only one thread may access a resource at any given time.
The resource is therefore guarantee exclusion of the resource from any thread whenever any other thread has access.

\subsection{Race Conditions}

O'Hearn begins by considering concurrent processes with access to shared, mutable state.
Because the state can be modified by each process, it is possible that they compete for access to resources, possibly clobbering computations.
Processes that attempt to access shared state at the same time are defined as \emph{racy}, and all other processes as \emph{race-free}.
He further points out that most formal frameworks make no claims about raciness, despite programmers understanding that the problem and its solutions are of great importance.

Next, processes that only access shared state within the associated mutual exclusion group are defined as \emph{cautious}, while all others are defined to be \emph{daring}.
Although cautious programs are typically well-behaved and easier to reason about, daring programs are very common.
As an example, the author cites Brinch Hansen who states that almost all semaphore programs are daring, as compilers cannot generally decide whether semaphore guards delimit critical regions, or if a missing guard is an error.
Daringness is often not an error but a part of the program's design, such as in processes that pass messages via reference into and out of a mutual exclusion group.

\subsection{Ownership and Separation}

Applying Separation Logic to concurrency requires the development of a few new notions: \emph{ownership} and \emph{separation}.

The \emph{Ownership Hypothesis} is stated to be
\begin{quote}
    A code fragment can access only those portions of state that it owns.
\end{quote}
This hypothesis is directly responsible for the \emph{Separation Property}, which states
\begin{quote}
    At any time, a state can be partitioned into that owned by each process and each grouping of mutual exclusion,
\end{quote}
clearly a descendant of Separation Logic's frame axiom.
It is noted that the Separation Property does not specify a static, unchanging partition, but one that can evolve as the program executes.
These statements are necessary to reason about independent program components, which eventually build up to include the entire program.
An example using semaphores is provided:

\begin{tabular}{c||c}
    $free \coloneqq 1$ & $busy \coloneqq 0$ \\
    \hline
    $P(free);$ & $P(busy);$ \\
    $[10] \coloneqq m;$ & $n \coloneqq [10];$ \\
    $V(busy);$ & $V(free);$
\end{tabular}

In this example, the author introduces the concept of resources being owned by semaphores, which can release or restrict these resources to or from the code that follows.
This code is treated as daring, because the resource at address 10 could be interacted with simultaneously if the $free$ and $busy$ semaphores were not instantiated correctly.
With the Ownership Hypothesis and Separation Property, we can state and prove assertions such as:
\begin{quote}
    address 10 is owned by exactly one of the left process, the right process, the free
semaphore or the busy semaphore.
\end{quote}
O'Hearn summarizes this development as allowing for the verification of daring systems if the transfer of state ownership can be reasoned about.

\subsection{Disjoint Processes}

Disjoint concurrent processes are considered as an introduction to the modification of Separation Logic for concurrency.
The axiom for disjoint programs is given:
\infrule{\triple{P}{C}{Q} & \triple{P'}{C'}{Q'}}{\triple{P * P'}{C || C'}{Q * Q'}}{}

O'Hearn provides an implementation of mergesort to show the reasoning capabilities of this rule:%
\begin{align*}%
    & \{array(a)\} \\
    & \texttt{procedure ms}(a, i, j) \\
    & \texttt{newvar } m \coloneqq (i + j) / 2 \texttt{;} \\
    & \texttt{if } i < j \texttt{ then} \\
    & \qquad (\texttt{ms}(a, i, m) || \texttt{ms}(a, m + 1, j)) \texttt{;} \\
    & \qquad \texttt{merge}(a, i, m + 1, j) \texttt{;} \\
    & \{sorted(a)\}
\end{align*}
A succinct proof is provided detailing how the provided axiom succeeds in verifying properties that involve passing the same variable to multiple subprocesses, an area in which previous Separation Logic extensions failed.

\subsection{Interacting Processes}

Interacting processes are presented as bound by \emph{conditional critical regions} (CCRs).
CCRs are a proposed lanugage construct that are chosen to generalize semaphores.
CCRs take the form of $\texttt{with } r \texttt{ when } B \texttt{ do } C$, ensuring that the program fragment $C$ executes atomically with exclusive access to resource $r$ whenever $B$ evaluates to true.
A set of proof rules are provided, describing the circumstances of resource initialization before concurrency, the composition of parallelism among many processes, and critical regions.

The author continues to provide examples of the versatility of the extended Separation Logic, including a binary semaphore, a pointer-transferring buffer, and a simple memory manager, with succinct and reasonable proofs of structural integrity and safety.

\subsection{Summary and Analysis}

O'Hearn develops a novel extension of Separation Logic to reason about concurrent programs, building on its success in reasoning about single-threaded programs.
The emphases on ownership and separation provide a natural segue from standard Separation Logic principles into concurrency, allowing for mass reuse of Separation Logic techniques.
Because of the generality of the underlying framework, O'Hearn's extension is capable of reasoning about arbitrary concurrent code, however some complex scenarios such as resource access permissions or network topology may be too complex to succinctly describe.

Although this extension loses no generality, it fails to reason about the \emph{liveness} of resources, a critical property of real-world systems.
Despite this significant missing piece of the theory, the multitude of examples tackling real-world scenarios ensures that the extended Separation Logic is extremely capable.

\section{Iris: Monoids and Invariants as an Orthogonal Basis for Concurrent Reasoning}

\subsection{Background}


This work proposes Iris, a further extension of Separation Logic beyond what O'Hearn proposes.
During the time between these two developments, a large number of separation logics were proposed to handle various issues with expressivity and simplicity of concurrent systems.
The key progress that Iris makes is generalizing many of the redundant or over-complicated components of these various logics, resulting in a consistent separation logic amenable to machine-checked verification of concurrent programs.

\subsection{Invariants}

The standard model of concurrency is \emph{sequential consistency}, in which threads take turns modifying and reading from shared memory.
This is a simple model to encode, but in practice is difficult to utilize for program verification.
A key area in which sequential consistency remains difficult to use is in \emph{thread-local} reasoning, in which each individual thread is verified to be correct, rather than the system as a whole.
This is difficult because threads in concurrent systems affect each others' correctness as shared data is modified.
The way in which thread-local reasoning is accomplished in this case utilizes \emph{invariants}, or properties of a partition of shared memory that are true before every program step, and must be re-proved after each program step.
\infrule{\triple{R * P}{c}{R * Q} & c \textrm{ is physically atomic}}{\textrm{invariant } R \textrm{ describes } \triple{P}{c}{Q}}{Invariant Rule}
Although elegant, the Invariant Rule is fairly primitive, suffering in two areas particularly.


\textbf{Invariants are too static}.
Real systems are dynamic and therefore have complex evolutions over time.
Therefore, it is often necessary to extend the logic of invariants into a system more comfortable with a changing state.
As a result, many separation logics bake in \emph{protocols}, descriptions of how the state changes over time, to the logic as primitives.
Although effective, the various implementations of these protocols seem to imply that there is a more generic pattern to be found.

\textbf{Atomicity is overreaching}.
Because the Invariant Rule requires atomicity of its code argument, it is not possible to apply it to many common scenarios.
For instance, implementations of concurrent data structures should shield complexity from the client code that uses them.
Even though a push to a concurrent stack might not be physically atomic, it is inconvenient to require a piece of code that utilizes concurrent stacks to consider the atomicity of the internals of the stack's implementation.
The authors explain that optimally, atomicity is contained within the language of Separation Logic, allowing for a more flexible and modular notion of \emph{logical atomicity}.
Other works have attempted this, but do not achieve the simplicity and ease of use the authors search for.

\subsection{Monoids and Iris}

A \emph{monoid} is a standard algebraic structure consisting of a set $M$, an operation $\cdot : M \rightarrow M \rightarrow M$ (compose), and an identity element $\epsilon$ such that the following properties are satisfied:
\begin{gather*}
    \textbf{Associativity } (a \cdot b) \cdot c = a \cdot (b \cdot c) \\
    \textbf{Identity } a \cdot \epsilon = \epsilon \cdot a = a
\end{gather*}
Some examples of common monoids are $(\mathbb{N}, +, 0)$ (the natural numbers under addition) and $(\texttt{string}, concat, ``")$ (strings under concatenation).

Monoids play a role in separation logic because \emph{heaps} form a monoid under heap concatenation.
As a primary contribution, the authors discuss how monoids enable Separation Logic to \emph{express} properties and protocols on a heap, while invariants enable the logic to \emph{enforce} these same properties and protocols.

To illustrate this discovery, the authors detail embeddings of common invariant protocols into Iris, derive a number of proof rules that other logics assume as axioms, encode logically-atomic specification rules, and verify a few concurrent abstract data types.
The \textbf{exclusive monoid} describes the availability of a heap, naturally encoding the concept of exclusive ownership.
The \textbf{fractional monoid} performs a kind of extrusion of knowledge about the \emph{ghost state}, an alternate program state that has no effect on program execution but simplifies the expression of high-level properties of the physical state.
The \textbf{product monoid} combines multiple monoids without losing any of their reasoning principles.
The \textbf{authoritative monoid} acts as an owner of some piece of shared ghost resource, while other parties can access fragments of the resource with the knowledge that the resource is contained in the authoritative state.

Logical atomicity is derived through another novel contribution: the \emph{view shift}.
The view shift is a logical transformation that permits readjustment of proof obligations in the logic.
These transformations are critical for making progress in goals regarding the open and close of invariants, the transferring of resource ownership, and manipulation of ghost state.

Iris is the result of extending standard Separation Logic with the aforementioned developments on invariants and monoids.
The authors have formalized the semantics of Iris in the Coq Proof Assistant, proving their soundness alongside other critical properties.

\subsection{Summary and Analysis}


This work presents Iris, a significant advancement in the field of separation logic that addresses critical limitations in existing frameworks for reasoning about concurrent programs. 
It extends traditional separation logic by introducing a systematic approach to invariants and monoids, resulting in a more expressive and modular system.
By leveraging the mathematical structure of monoids, Iris provides a unified framework to encode and enforce resource protocols.
The integration of logical atomicity through view shifts further enhances the framework, enabling it to handle complex reasoning tasks such as dynamic state changes and the encapsulation of atomicity.
These features make Iris a versatile tool for verifying concurrent systems, particularly those requiring machine-checked proofs.

Unfortunately, Iris has reached critical levels of complexity that makes the system difficult to learn and use when compared to the relatively restricted Hoare and Separation logics.
Additionally, this work fails to provide examples of application of Iris to real-world scenarios.
Despite these fallbacks, Iris is an incredibly general system, giving it the power to represent arbitrary systems.


\section{Evaluation}

Now we compare and contrast the four works along three critical axes: \emph{generality} --- how many kinds of concurrent systems a framework can represent, \emph{simplicity} --- how easy it is to utilize a framework for analysis of concurrent programs, and the \emph{impact} that each work had on the field.

\subsection{Generality}

Each of the four provided frameworks can encode information about arbitrary concurrent systems, but not all do so effectively.
Hoare's extension of the Hoare logic conveys only qualitatively the extent to which it can be used on generic systems.
The extension largely focuses on thread-local verification, but fails to detail how complex, evolving systems could be represented.
In direct contrast, Milner's \picalc\ represents evolving states with extreme clarity and simplicity through scope extrusion and the generic concept of link passing.
O'Hearn's development on Separation Logic falls into similar patterns seen in Hoare's work and fails to prove its generality, but indeed accomplishes its goal of succinctly generalizing the concepts of ownership and separation across concurrent processes.
Finally, Iris is designed to be completely general and manages to accomplish this with a clear argument that it can represent arbitrary systems.

\subsection{Simplicity}

Hoare's development is by far the simplest - providing a small set of extensions on the standard Hoare logic for reasoning about concurrency.
This results in a system that is incredibly simple to use, requiring very few alterations to a standard formalization of Hoare logic and little extra study for verifying the properties of concurrent systems.
The generality of the \picalc\ is its downfall in simplicity.
Although it is tasked with encoding the behaviors of complex, evolving systems, the \picalc\ and its semantics are difficult to gain intuitive understanding of without sufficient study and analysis.
O'Hearn's extension of Separation Logic retains the vast simplicity of Hoare's earlier work, again requiring only a small amount of extension work required to update a Separation Logic system to support concurrency.
Finally, Iris is the most complex out of each of these systems, partially as a result of being tasked with being formalized from the start.
It requires large amounts of study and practice to become familiar with the system in order to use it for real-world verification.

\subsection{Impact}

Hoare's work is in every way visionary --- predicting the importance of exclusion principles and in general laying out the foundation for what would eventually become Separation Logic, even before it was known to be useful for concurrent systems.
It is in every way a cornerstone of the field of formal concurrency.
The \picalc\ has similarly become vastly important to the development of domain-specific languages for multiprocessing and networking, allowing compilers to automatically analyze the behavior of complex concurrent systems.
This work is frequently taught in universities as a foundation for complex parallelism.
O'Hearn's extension of Separation Logic to concurrency was a massive development, spawning off countless more extensions to support various concurrency features.
Finally, although Iris is much newer in comparison with the other three works, it has become a standard in the world of mechanized formal verification due to its generality and power.


\bibliographystyle{acm}
\bibliography{sample}

\end{document}